\documentclass[aps,showpacs,manuscript,12pt]{revtex4}
\usepackage{amssymb}
\usepackage{amsmath}
\usepackage{graphicx}

\begin{document}
\title{\textbf{Fokker-Planck Kinetic description of small-scale fluid turbulence for
classical incompressible fluids$^{\S }$}}
\author{M. Tessarotto$^{a,b}$, M. Ellero$^{c}$, D. Sarmah$^{b}$ and P. Nicolini$^{a,b}$}
\affiliation{\ $^{a}$Department of Mathematics
and Informatics, University of Trieste, Italy, $^{b}$Consortium of
Magneto-fluid-dynamics, University of Trieste, Italy,
$^{c}$Department of Aerodynamics, Technical University of Munich,
Munich, Germany}
\begin{abstract}
Extending the statistical approach proposed in a parallel paper \cite%
{Tessarotto2008-aa}, purpose of this work is to propose a
stochastic inverse kinetic theory for small-scale hydrodynamic
turbulence based on the introduction of a suitable \textit{local
phase-space probability density function} (pdf). In particular, we
pose the problem of the construction of Fokker-Planck kinetic
models of hydrodynamic turbulence. The approach here adopted is
based on the so-called IKT approach (inverse kinetic theory),
developed by Ellero et al. (2004-2008) which permits an exact
phase-space description of incompressible fluids based on the
adoption of a local pdf. We intend to show that for prescribed
models of stochasticity the present approach permits to determine
uniquely the time evolution of the stochastic fluid fields. The
stochastic-averaged local pdf is shown to obey a kinetic equation
which, although generally non-Markovian, locally in velocity-space
can be approximated by means of a suitable Fokker-planck kinetic
equation. As a side result, the same pdf is proven to have
generally a non-Gaussian behavior.
\end{abstract}
\maketitle



\section{Introduction}

In this paper - extending the results of
Ref.\cite{Tessarotto2008-aa} - we
intend to formulate an\emph{\ IKT} (inverse kinetic theory) approach \emph{%
for the full set of fluid equations describing the phenomenon of
turbulence in an incompressible fluid}, here identified with the
stochastic
incompressible Navier-Stokes equations. Unlike Ref. \cite{Tessarotto2008-aa}%
, where the set of stochastic-averaged fluid equations were
considered, we intend to show that the kinetic description, for
the stochastic local probability distribution function
(\emph{stochastic local pdf}), is generally non-Markovian when the
complete set of fluid equations is considered. The theory is shown
to satisfy an H-theorem assuring the strict positivity of the
local pdf as well of its stochastic average. In particular, the
stochastic-averaged pdf is shown to satisfy an approximate
Fokker-Planck-type kinetic equation. The explicit representation
of the leading Fokker-Planck coefficients is provided. The result,
which holds in principle for arbitrary prescribed stochasticity of
the fluid fields, is achieved by means of an IKT which delivers
the complete set of fluid fields, all expressed in terms of
appropriate moments of the stochastic local pdf.

\subsection{Background and open problems}

The investigation \emph{hydrodynamic turbulence} in incompressible
fluids is nowadays playing a major role in fluid dynamics
research. In fact the phenomenon of turbulence is essentially
ubiquitous,\ being related to its statistical character. Indeed in
many cases the fluid fields which define an
incompressible isothermal fluid, i.e., the fluid velocity and pressure ${Z(%
\mathbf{r,}t\mathbb{)}\equiv }\left\{ \mathbf{V},p\right\} $ are
actually not known deterministically but only in a statistical
sense. This means that
the fluid fields must contain some kind of parameter-dependence $Z(\mathbf{r}%
,t,\mathbf{\alpha })$, where $\mathbf{\alpha \equiv }\left( \alpha
_{1},...,\alpha _{n}\right) \in V_{\alpha }\subseteq
\mathbb{R}
^{n}$ denotes a suitable stochastic real vector independent of ${(\mathbf{r,}%
t\mathbb{)}}$ to which a \emph{stochastic probability density}
$g(\alpha )$
can be attached, so that $\int\limits_{V_{\alpha }}d^{n}\mathbf{\alpha }g(%
\mathbf{\alpha })=1$ and furthermore the \emph{stochastic average }$%
\left\langle Z(\mathbf{r},t,\alpha )\right\rangle \equiv
\int\limits_{V_{\alpha }}d^{n}\alpha g(\alpha
)Z(\mathbf{r},t,\alpha )$ exists. As a consequence, the fluid
fields can be represented in terms of
the \emph{stochastic decomposition }$Z(\mathbf{r,}t,\alpha )=\left\langle Z(%
\mathbf{r,}t)\right\rangle +\delta Z(\mathbf{r,}t,\alpha ),$ where $\delta {%
Z(\mathbf{r,}t,\alpha \mathbb{)\equiv }}\left\{ \delta
\mathbf{V},\delta p\right\} $ are suitable \emph{stochastic
fluctuations} of the fluid fields. \ The precise form which these
fluctuations may take defines what is usually denoted as the
\emph{stochastic behavior} (or \emph{stochasticity}) of the fluid.
\ The vector $\alpha ,$ which spans a suitable subset $V_{\alpha
}$
of $%
\mathbb{R}
^{n},$ can in principle be assumed either continuous or discrete.
\ Its possible definition, as well the identification of the
related probability density appearing in the stochastic-averaging
operator $\left\langle {}\right\rangle $, \ is however manifestly
non-unique. In fact, these definitions are closely related to the
types of stochastic behavior which may appear in the fluid. In the
mathematical theory of turbulence one can
distinguish in principle two possible types of stochasticity: either \emph{%
intrinsic} or \emph{numerical}. The first type (\emph{intrinsic stochasticity%
}) arises when, \emph{leaving unchanged the functional form of the
fluid equations}, the fluid equations are intended as stochastic
pde's. This happens if\ at least one of the following
\emph{sources of stochasticity} is introduced: 1) \emph{Stochastic
initial conditions:} in this case the initial fluid fields
$Z(\mathbf{r,}t_{o})\equiv Z_{o}(\mathbf{r})$ are assumed
stochastic, i.e. of the form, $Z_{o}(\mathbf{r},\alpha
)=\left\langle Z_{o}(\mathbf{r},\alpha )\right\rangle +\delta Z_{o}(\mathbf{r%
},\alpha ),$ being $\left\langle Z_{o}(\mathbf{r},\alpha
)\right\rangle $
and $\delta Z_{o}(\mathbf{r},\alpha )$ suitable vector fields. 2) \emph{%
Stochastic boundary conditions:} this happens if the boundary fluid fields $%
\left. Z_{w}(\mathbf{r,}t)\right\vert _{\delta \Omega }$ are
prescribed in
terms of a suitable stochastic vector field of the form $\left. Z_{w}(%
\mathbf{r,}t,\alpha )\right\vert _{\delta \Omega }=\left. \left\langle Z_{w}(%
\mathbf{r,}t,\alpha )\right\rangle \right\vert _{\delta \Omega
}+\left. \delta Z_{w}(\mathbf{r,}t,\alpha )\right\vert _{\delta
\Omega }.$ Here, Dirichlet (no-slip) boundary conditions have been
imposed (for the fluid fields) on the boundary set $\delta \Omega
$ of the fluid domain $\Omega \subseteq \mathbb{R}^{3}$ by letting
$\left. Z(\mathbf{r,}t)\right\vert _{\delta \Omega }=\left.
Z_{w}(\mathbf{r,}t)\right\vert _{\delta \Omega }.$ 3)
\emph{Stochastic forcing.} In this case the volume force density
acting
on the fluid is assumed stochastic, i.e., of the form $\mathbf{f}(\mathbf{r,}%
t,\alpha )=\left\langle \mathbf{f}(\mathbf{r,}t,\alpha
)\right\rangle
+\delta \mathbf{f}(\mathbf{r,}t,\alpha ),$ being $\left\langle \mathbf{f}(%
\mathbf{r,}t,\alpha )\right\rangle $ and $\delta \mathbf{f}(\mathbf{r,}%
t,\alpha )$ suitable vector fields. It is obvious that the parameters $%
\mathbf{\alpha \equiv }\left( \alpha _{1},...,\alpha _{n}\right) $
and the related probability density $g(\mathbf{\alpha })$ can be
set, in principle, arbitrarily. In fact, no information on them
can be gathered from the deterministic fluid equations. \ For the
same reason, also the definition of
the stochastic fluctuations appearing in the previous equations, namely $%
\delta Z_{o}(\mathbf{r},\mathbf{\alpha }),$ $\left. \delta Z_{w}(\mathbf{r,}%
t,\mathbf{\alpha })\right\vert _{\delta \Omega }$ and $\delta \mathbf{f}(%
\mathbf{r,}t,\mathbf{\alpha }),$ remains unspecified. \ As a
consequence,
each of the stochastic vector fields $Z_{o}(\mathbf{r},\mathbf{\alpha }),$ $%
\left. Z_{w}(\mathbf{r,}t,\mathbf{\alpha })\right\vert _{\delta
\Omega }$ and $\mathbf{f}(\mathbf{r,}t,\mathbf{\alpha })$ may in
principle be characterized by different stochastic parameters
$\mathbf{\alpha }$ and probability densities $g(\mathbf{\alpha
})$. \ Regarding, in particular, the definition of the boundary
conditions, we remark that the stochasticity of the boundary fluid
fields $\left. Z_{w}(\mathbf{r,}t,\alpha )\right\vert _{\delta
\Omega }$ may be simply a result of the choice of the prescribed
boundary $\delta \Omega .$ This happens if $\delta \Omega $ is
identified, for example, with a moving surface in which each point
(of the surface) moves with random motion. The second type
(\emph{numerical stochasticity}) arises - instead - as a result of
\ the approximate numerical solution methods adopted. \ All
numerical methods, in fact, involve in some sense the introduction
of appropriate approximations for the relevant differential
operators, based on suitable time and space discretizations. As a
consequence, the numerical solutions obtained for the fluid fields
become inaccurate on scale lengths comparable or smaller than the
spatial discretization (or grid) scales, thus producing stochastic
error fields. \ As a result, even if the fluid equations are
treated as deterministic (which means ignoring all possible
sources of intrinsic turbulence indicated above), numerical errors
produce fluid fields which behave effectively as stochastic,
giving rise to the phenomenon of numerical turbulence, also known
as \emph{small-scale }or \emph{sub-grid turbulence.} This means
that in principle it is possible to treat the two problems in a
similar way. However, numerical stochasticity, as opposite to
intrinsic stochasticity, is expected - in principle - to allow a
well-defined statistical description. Nevertheless, a consistent
theoretical formulation of small-scale turbulence of general
validity seems still far away. In particular, still missing is a
consistent statistical description, based on the proper definition
of the stochastic \emph{\ }probability density $g(\alpha )$, \
able to recover from first principles the correct form of the pdf
appropriate for arbitrary turbulence regimes. A widespread picture
(of turbulence) consists both of an ensemble of finite-amplitude
waves with random phase. However, there is an increasing evidence
that this picture is an oversimplification. In fact, it is well
known that turbulence may include fluctuations whose
phase-coherence characteristics are incompatible with wave-like
properties. These are so-called coherent structures, like shocks,
vortices and convective cells. In fluid turbulence the signature
of the presence of coherent structures is provided by the
existence of non-Gaussian features in the probability density.
This is usually identified with the velocity-difference
probability density function (pdf), traditionally adopted for the
description of homogeneous turbulence. \ This explains why in the
past the treatment of hydrodynamic turbulence was based on
stochastic models of various nature. These models, which are based
on tools borrowed from the study of random dynamical systems,
typically rely - however - on experimental verification rather
than on first principles. An example is provided by stochastic
models - based on Markovian Fokker-Planck (F-P) models of
small-scale fluid turbulence recently investigated in the
literature by several authors (including: Naert \textit{et al.},
1997 \cite{Naert1997}; Friedrich and Peinke \textit{et al.}, 1999
\cite{Friedrich1999}; Luck \textit{et al.}, 1999 \cite{Luck1999};
Cleve \textit{et al.}, 2000 \cite{Cleve2000}; Ragwitz
and Kantz, 2001 \cite{Ragwitz2001}; Renner \textit{et al., }2001, 2002 \cite%
{Renner2001,Renner2002}; Hosokawa, 2002 \cite{Hosokawa2002}). The
validity of phenomenological statistical Markovian Fokker-Planck
(F-P) models of small-scale fluid turbulence indicate that they
are capable of reproducing correctly, at least in some approximate
sense, key features of the basic phenomenology of turbulent flows.
Their approach is based on the assumption that the probability
density associated to the velocity increments should
obey a stationary generalized F-P equation. Experimental evidence \cite%
{Naert1997} shows reasonable agreement both with the Markovian
assumption and the F-P approximation, at least in a limited subset
of parameter space. However, several aspects of the theory need
further investigations. In particular, still missing is a
consistent statistical description following uniquely from the
fluid equations. The theory should be able, specifically, to
recover correctly the structure functions, characteristic for the
appropriate turbulence regime, but also - possibly - to apply to
non-Gaussian probability densities. The latter is, in fact, a
typical feature suggested by experimental observations, performed
at sufficiently short scale-lengths in the inertial
range\cite{Falkovich2001}. Based on a recently proposed inverse
kinetic theory for classical and quantum fluids (Ellero and
Tessarotto, 2004-2008 \cite{Ellero2005,Tessarotto2006}), a
statistical model of small-scale hydrodynamic turbulence is
proposed which holds for a generic form of the stochastic
probability density $g(\alpha )$ . The approach is intended to
determine the local pdf's (i.e., the \underline{local}
position-velocity joint probability density functions),
both for the stochastic-averaged fluid fields $\left\langle Z(\mathbf{r,}%
t)\right\rangle $ and - unlike Ref. \cite{Tessarotto2008-aa} -
also their stochastic fluctuations $\delta Z(\mathbf{r,}t,\alpha
),$ respectively denoted as $\left\langle f\right\rangle $ and
$\delta f$. In particular, it is proven that $\left\langle
f\right\rangle $ and $\delta f$ uniquely
determine, by means of suitable velocity-moments, the fluid fields $%
\left\langle Z(\mathbf{r,}t)\right\rangle $ and $\delta Z(\mathbf{r,}%
t,\alpha )$. Key feature of the approach concerns the construction
of the
statistical evolution equations for $\left\langle f\right\rangle $ and $%
\delta f$. In particular, it is shown that $\left\langle
f\right\rangle $ is generally non-Gaussian and obeys an H-theorem.
Finally, $\left\langle f\right\rangle $ it is shown to obey -
under suitable asymptotic assumptions - to an approximate
Fokker-Planck kinetic equation which hold in principle even in the
case on non-stationary, non-isotropic and non-homogenous
turbulence.

\section{Stochastic INSE and stochastic IKT}

It is convenient first to recall the basic equations for the
average and stochastic fluid fields and the corresponding
initial-boundary value problem. Starting from the incompressible
Navier-Stokes (NS) equations (INSE) and invoking the stochastic
decomposition given above the relevant
stochastic fluid equations read%
\begin{eqnarray}
&&\left. \left\langle N\mathbf{V}\right\rangle =\mathbf{0}\right.
\label{average NS} \\
&&\left. \nabla \cdot \left\langle \mathbf{V}\right\rangle
=0\right. , \label{physical realiz.2-average}
\end{eqnarray}%
\begin{eqnarray}
&&\left. N\mathbf{V}-\left\langle N\mathbf{V}\right\rangle =\mathbf{0,}%
\right.  \label{stoch. INSE} \\
&&\left. \nabla \cdot \delta \mathbf{V}=0,\right. \label{physical
realiz. 2 - stoch}
\end{eqnarray}%
to be denoted as stochastic incompressible NS equations
(\emph{stochastic
INSE}) and $N$ is the Navier-Stokes operator $N\mathbf{V=}\frac{\partial }{%
\partial t}\mathbf{V}+\mathbf{V\cdot \nabla V}+\frac{1}{\rho _{o}}\left[
\mathbf{\nabla }p-\mathbf{f}\right] -\nu \nabla ^{2}\mathbf{V.}$
In
particular the first ones (\ref{average NS})-(\ref{physical realiz.2-average}%
) are hereon denoted as \emph{stochastic-averaged INSE}. These
equations are assumed to be satisfied pointwise in a set $\Omega
\times I,$ being $\Omega $
an open subset of $%
\mathbb{R}
^{3}$ and $I$ a finite time interval. Assuming that the fluid fields and $%
\mathbf{f}$ are sufficiently smooth, the conditions of
isochoricity and incompressibility imply the validity of the
Poisson equations respectively for $\left\langle p\right\rangle $
and $\delta p.$ Let us now adopt the statistical approach
developed in Ref.\cite{Ellero2005}, which allows us to cast the
stochastic INSE problem in terms of a so-called inverse kinetic
theory (IKT) \cite{Tessarotto2006}. This is based on the the
identification of the fluid fields $Z=\left\{ \mathbf{V},p\right\}
$ with the moments of a suitable probability density
$f(\mathbf{x},t;Z)$ defined in the extended
phase space $\Gamma \times I$ [being $\mathbf{x}$ the state vector $\mathbf{x%
}=(\mathbf{r,v})\in \Gamma $ spanning the phase-space $\Gamma
=\Omega \times
V,$ with $\Omega $ the fluid domain and $V=%
\mathbb{R}
^{3}$ the corresponding velocity space]. In particular, it follows that $f(%
\mathbf{x},t;Z)$ must obey an inverse kinetic equation
(\emph{IKE}), which can be identified with the Vlasov-type
equation

\begin{equation}
L(Z)f=0.  \label{inverse kinetic equation}
\end{equation}%
Here the notation is standard. Thus $L$ is the streaming operator $%
L(Z)\equiv \frac{\partial }{\partial t}+\frac{\partial }{\partial \mathbf{x}}%
\cdot \left\{ \mathbf{X}(Z)\right\} $ and $\mathbf{X}(Z)=\left\{ \mathbf{v,F}%
(Z)\right\} ,$ with $\mathbf{F}(Z)\equiv
\mathbf{F}(\mathbf{x},t;f,Z)$ a suitable vector field (mean field
force). Subject only to the requirement of suitable smoothness for
the fluid fields, several important consequences follow
\cite{Ellero2005}, in particular:

\begin{itemize}
\item The mean-field force reads:%
\begin{equation*}
\mathbf{F}(\mathbf{x},t;f,Z)=\frac{1}{\rho _{o}}\left[ \mathbf{\nabla \cdot }%
\underline{\underline{\Pi }}-\mathbf{\nabla }p_{1}-\mathbf{f}\right] +\frac{1%
}{2}\mathbf{u}\cdot \nabla \mathbf{V+}\frac{1}{2}\mathbb{\nabla }\mathbf{%
V\cdot u}+\nu \nabla ^{2}\mathbf{V}+
\end{equation*}%
\begin{equation}
+\frac{1}{2}\mathbf{u}\left[ \frac{\partial \ln p_{1}}{\partial t}+\mathbf{%
V\cdot \nabla }\ln p_{1}+\frac{1}{p_{1}}\left( \nabla \cdot
\mathbf{Q-}\left[
\nabla \cdot \underline{\underline{{\Pi }}}\right] \cdot \mathbf{\mathbf{Q}}%
\right) \right] +\frac{1}{\rho _{o}}\mathbf{\nabla \cdot }\underline{%
\underline{\Pi }}\left\{ \frac{\mathbf{u}^{2}}{v_{th}^{2}}-\frac{3}{2}%
\right\} ,
\end{equation}
\end{itemize}

where $\mathbf{Q}$ and $\underline{\underline{{\Pi }}}$ are the
velocity
moments $\int d^{3}vGf$ for $G=\mathbf{u}\frac{u^{2}}{3},\mathbf{uu}$ and $%
\mathbf{u}\mathbb{\equiv }\mathbf{v}-\mathbf{V}(\mathbf{r,}t)$ is
the relative velocity$.$

\begin{itemize}
\item $\left\{ \mathbf{V,}p\right\} $can be identified in the whole fluid
domain $\Omega $ with the velocity moments
$G=\mathbf{v,}\frac{1}{3}u^{2}$\ of $f(\mathbf{x},t;Z),$i.e.,
respectively, $\mathbf{V}(\mathbf{r,}t)=\int
d^{3}v\mathbf{v}f(\mathbf{x},t;Z)$ and $p\mathbf{(r,}t)=p_{1}\mathbf{(r,}%
t)-P_{0},$ where $P_{0}(t)$ (pseudo-pressure) is an arbitrary
strictly positive real function of time defined so that the
physical realizability
condition $p\mathbf{(r,}t)\geq 0$ is satisfied everywhere in $\overline{%
\Omega }\times I$ and $p_{1}(\mathbf{r,}t)=\rho _{o}\int d\mathbf{v}\frac{1}{%
3}u^{2}f(\mathbf{x},t;Z)$ is the kinetic scalar pressure.

\item If $f(\mathbf{x},t;Z)$ is a strictly positive and summable
phase-function in $\Gamma ,$ the Shannon entropy functional\ $%
S(f(t))=-\int_{\Gamma }d\mathbf{x}f(\mathbf{x},t;Z)\ln
f(\mathbf{x},t;Z)$ exists $\forall t\in $ $I$ and can be required
to fulfill a constant
H-Theorem (see Ref.\cite{Tessarotto2008}), i.e., $\frac{\partial }{\partial t%
}S(f(t))=0.$

\item Introducing the notation $x^{2}=\frac{\mathbf{u}^{2}}{v_{th}{}^{2}},$ $%
v_{th}^{2}=2p_{1}/\rho _{o},$ the pdf
\begin{equation}
f_{M}(\mathbf{x,}t;Z)=\frac{\rho _{o}^{3/2}}{\left( 2\pi \right) ^{\frac{3}{2%
}}p_{1}^{\frac{3}{2}}}\exp \left\{ -x^{2}\right\}
\label{Maxwellian}
\end{equation}
(\emph{local Maxwellian kinetic equilibrium}) is a particular
solution of the inverse kinetic equation (\ref{inverse kinetic
equation}) if and only if $\left\{ \mathbf{V,}p\right\} $ satisfy
INSE.
\end{itemize}

It is now immediate to obtain an inverse kinetic theory for the
previous
stochastic fluid equations. In fact, let us assume that the operator $%
\left\langle {}\right\rangle $ is taken at constant
$\mathbf{r},\mathbf{v}$
and $t$ and introduce the stochastic decompositions $f(\mathbf{x},t;Z)%
\mathbb{=}\left\langle f(\mathbf{x},t;Z)\right\rangle +\delta f(\mathbf{x}%
,t;Z),$ $L(Z)\mathbb{=}\left\langle L(Z)\right\rangle +\delta L(Z)$ and $%
\mathbf{F}(\mathbf{x},t;f,Z)=\left\langle \mathbf{F}\right\rangle
+\delta
\mathbf{F,}$ where $\left\langle L(Z)\right\rangle =\frac{\partial }{%
\partial t}+\mathbf{v\cdot }\frac{\partial }{\partial \mathbf{r}}+\frac{%
\partial }{\partial \mathbf{v}}\cdot \left\{ \left\langle \mathbf{F}%
(Z)\right\rangle \right\} $ and respectively $\delta L(Z)=\frac{\partial }{%
\partial \mathbf{v}}\cdot \left\{ \left\langle \mathbf{F}(Z)\right\rangle
\right\} .$ Then the following theorem holds:\newline

\textbf{Theorem 1 - Stochastic IKT for the stochastic INSE
problem}\newline
\emph{If }$f(\mathbf{x},t;Z)$ \emph{is a particular solution of the IKE [Eq.(%
\ref{inverse kinetic equation})] then it follows that:} \emph{B1)} $%
\left\langle f(\mathbf{x},t;Z)\right\rangle $ \emph{and }$\delta f(\mathbf{x}%
,t;Z)$ \emph{obey the coupled system of} \emph{stochastic kinetic equations}%
\begin{equation}
\left\langle L(Z)\right\rangle \left\langle f\right\rangle
=-\left\langle \delta L(Z)\delta f\right\rangle \equiv
\left\langle C\right\rangle , \label{stochasrtic-averaged IKE}
\end{equation}%
\begin{equation}
\left\langle L(Z)\right\rangle \delta f=-\delta L(Z)\left\{
\left\langle f\right\rangle +\delta f\right\} +\left\langle \delta
L(Z)\delta f\right\rangle ;  \label{stochastic IKE}
\end{equation}
\emph{B2) A particular solution is provided by }$\left\langle
f\right\rangle =\left\langle f_{M}(\mathbf{x,}t;Z)\right\rangle
,$\emph{\ }$\delta f=\delta f_{M}(\mathbf{x,}t;Z);$ \emph{B3) Eq.
(\ref{stochasrtic-averaged IKE}) can also be written in the form}
\begin{equation}
\left\langle L(Z)\right\rangle \left\langle \Delta f(\mathbf{x}%
,t;Z)\right\rangle +\left\langle \Delta L\right\rangle f(\mathbf{x}%
,t;\left\langle Z\right\rangle )=-\left\langle \delta \Delta
L\delta \Delta f(\mathbf{x},t;\left\langle Z\right\rangle
)\right\rangle ,
\end{equation}%
\emph{where }$\Delta L=L(Z)$ $-L(\left\langle Z\right\rangle ),$ $\Delta f(%
\mathbf{x},t;Z)=f(\mathbf{x},t;Z)-f(\mathbf{x},t;\left\langle
Z\right\rangle
)$ \emph{and} $L(\left\langle Z\right\rangle ),$ $f(\mathbf{x}%
,t;\left\langle Z\right\rangle )$ \emph{are respectively the
stochastic-averaged streaming operator and pdf given in Ref.\cite%
{Tessarotto2008-aa}) [see Eq.(5)] and there results} $\left\langle
C\right\rangle =-\left\langle \delta \Delta L\delta \Delta f(\mathbf{x}%
,t;\left\langle Z\right\rangle )\right\rangle ;$ \emph{B4) If
}$P_{0}(t)$
\emph{is defined so that the constant H-theorem }$\frac{\partial }{\partial t%
}S(f(t))=0$\emph{\ is fulfilled identically in }$I,$ \emph{it follows that} $%
\left\langle f(\mathbf{x,}t;Z)\right\rangle $ \emph{satisfies the
weak H-theorem }$\frac{\partial }{\partial t}S(\left\langle
f(t)\right\rangle
)\geq 0,$\emph{\ hence both }$f(\mathbf{x,}t;Z)$ \emph{and} $\left\langle f(%
\mathbf{x,}t;Z)\right\rangle $ \emph{are} \emph{probability densities.}%
\newline
PROOF - The proof is immediate. In fact B$_{1}$) follows invoking
the inverse kinetic Eq.(\ref{inverse kinetic equation}) and the
stochastic
decompositions for $f(\mathbf{x},t;Z),L(\mathbf{F})$ and $\mathbf{F}(\mathbf{%
x},t;f,Z)$. B$_{2}$) is manifestly fulfilled since by construction $f_{M}(%
\mathbf{x},t;Z)$ is a particular solution of the inverse kinetic equation (%
\ref{inverse kinetic equation})$.$ Proposition B$_{3}$) follows by
noting
that $f(\mathbf{x},t;Z)$ can be represented as $f(\mathbf{x},t;Z)=f(\mathbf{x%
},t;\left\langle Z\right\rangle )+\Delta f(\mathbf{x},t;Z),$ where $f(%
\mathbf{x},t;\left\langle Z\right\rangle )$ is the
stochastic-averaged pdf
solution of the IKE of the form $\left. L(\left\langle Z\right\rangle )f(%
\mathbf{x},t;\left\langle Z\right\rangle )=0\right. ,$ where
$L(\left\langle
Z\right\rangle )$ is a suitable streaming operator (see Eq.(5) in Ref.\cite%
{Tessarotto2008-aa}). Finally for B$_{4}$) we notice that by construction $%
\int_{\Gamma }d\mathbf{x}\left\langle
f(\mathbf{x},t;Z)\right\rangle =1.$ It
follows $\int_{\Gamma }d\mathbf{x}f(\mathbf{x},t;Z)\ln \left\langle f(%
\mathbf{x},t;Z)\right\rangle \leq \int_{\Gamma }d\mathbf{x}f(\mathbf{x}%
,t;Z)\ln f(\mathbf{x},t;Z)$ and hence $S(f(t))\leq S(\left\langle
f(t)\right\rangle ).$ Hence, requiring since $P_{0}(t)$ can be
determined so that $\frac{\partial }{\partial t}S(f(t))=0$
\cite{Tessarotto2008}, this proves the H-theorem.

An obvious implication (of this result) is the manifest
non-Gaussian
behavior of the stochastic-averaged pdf $\left\langle f(\mathbf{x}%
,t;Z)\right\rangle .$ In fact, even if $f(\mathbf{x},t;Z)$ is
identified with the local Maxwellian distribution
$f_{M}(\mathbf{x,}t;Z),$ for arbitrary prescribed choices of the
stochastic probability density $g(\alpha )$ its stochastic average
is generally non-Gaussian since $\left\langle
f_{M}(\mathbf{x,}t;Z)\right\rangle \neq
f_{M}(\mathbf{x,}t;\left\langle
Z\right\rangle )\equiv \frac{\rho _{o}^{3/2}}{\left( 2\pi \right) ^{\frac{3}{%
2}}<p_{1}>^{\frac{3}{2}}}\exp \left\{ -\frac{<\mathbf{u>}^{2}\rho _{o}}{%
2<p_{1}>}\right\} $.

\section{Fokker--Planck approximation}

It is interesting to stress that Eqs. (\ref{stochasrtic-averaged IKE}) and (%
\ref{stochastic IKE}) are formally similar to the Vlasov equation
arising in the kinetic theory of quasi-linear and strong
turbulence for Vlasov-Poisson plasmas
\cite{Dupree1966,Weinstock1969,Benford1972} and related
renormalized kinetic theory \cite{Krommes1979}, which are known to
lead generally to a non-Markovian kinetic equation for
$\left\langle f\right\rangle $ alone.
Nevertheless, the stochastic-averaged kinetic equation [i.e., Eq. (\ref%
{stochasrtic-averaged IKE})] is known to be amenable, under
suitable assumptions, to an approximate Fokker-Planck kinetic
equation advancing in time $\left\langle f\right\rangle $ alone.
This is achieved by formally constructing a perturbative solution
of the equation (\ref{stochastic IKE}) for the stochastic
perturbation $\delta f.$ To obtain a convergent perturbative
theory, however, this usually requires the adoption of a suitable
renormalization scheme in order to obtain a consistent kinetic
equation for $\left\langle f\right\rangle .$ An analogous
suggestion is posed by the phenomenological Fokker-Planck models
of small-scale fluid turbulence recently investigated in the
literature. This suggests to seek for a possible approximate
representation of this type holding for the stochastic-averaged
kinetic equation (\ref{stochasrtic-averaged IKE}) which should
hold for generally non-Gaussian pdf's (in fact it is obvious that
generally the average distribution function $\left\langle f(\mathbf{x}%
,t;Z)\right\rangle ,$ even in the case in which it coincides with $%
\left\langle f_{M}(\mathbf{x},t;Z)\right\rangle $ results
generally non-Gaussian. Let us now first assume that $\left\langle
f\right\rangle \equiv $ $\left\langle
f_{M}(\mathbf{x},t;Z)\right\rangle .$ In such a case
the pseudo-pressure $P_{0}$ can be defined - consistent with Eq.(\ref%
{Fokker-Planck operator}) - so that locally in phase-space the
following asymptotic orderings
\begin{eqnarray}
&&\left. \frac{p}{P_{0}}\ll 1\sim o(\zeta ),\right.  \label{finite
amplitude}
\\
&&\left. X\equiv o(\zeta ),\right.  \label{finite-amplitude 2}
\end{eqnarray}%
are satisfied, being $\zeta $ a dimensionless infinitesimal
parameter. \ In such a case, without additional assumptions on the
amplitude of the
stochastic fields and in the sub-domain of velocity space in which (\ref%
{finite-amplitude 2}) holds, the "collision operator"
$\left\langle C\right\rangle $ in Eq.(\ref{average NS}) can be
approximated by the generalized Fokker-Planck (F-P) operator of
the form
\begin{equation}
\left\langle C\right\rangle \cong C_{FP}\equiv \sum \frac{\partial }{%
\partial \mathbf{v}}\cdot \left[ \mathbf{C}_{j+1,m}\cdot \frac{\partial ^{n}%
}{\partial ^{j}\mathbf{v}\partial
^{m}P_{0}}f(\mathbf{x},t;\left\langle Z\right\rangle )\right]
\label{Fokker-Planck operator}
\end{equation}%
with summation carried out on $j,m$ from $0$ to $\infty $ or to finite $%
N\geq 3$ and with $n=j+m>1,$ being $C_{FP}$ a F-P operator and $\mathbf{C}%
_{j,m}$ suitable F-P coefficients$.$ In the previous equation $f(\mathbf{x}%
,t;\left\langle Z\right\rangle )\equiv
f_{M}(\mathbf{x},t;\left\langle
Z\right\rangle )$ which denotes the local averaged-Maxwellian $f_{M}(\mathbf{%
x},t;\left\langle Z\right\rangle )=\frac{1}{\left( \pi \right) ^{\frac{3}{2}%
}v_{th}^{3}}\exp \left\{ -\widehat{X}^{2}\right\} ,$ with $\widehat{X}^{2}=%
\frac{\left\langle \mathbf{u}\right\rangle ^{2}}{vth^{2}},$ $%
v_{th}^{2}=2\left\langle p_{1}\right\rangle /\rho _{o}$ and
$\left\langle
\mathbf{u}\right\rangle \mathbf{=v}-\left\langle \mathbf{V}(\mathbf{r}%
,t)\right\rangle $. \ An analogous result holds also in the case $f\neq $ $%
f_{M}(\mathbf{x},t;Z).$ In particular, the following theorem
holds: \newline
\newline
\textbf{Theorem 2 - Approximate Fokker-Planck IKE }\newline
\emph{Let us assume that }$f(\mathbf{x},t;Z)$ \emph{is a
particular solution
of the IKE [Eq.(\ref{inverse kinetic equation})], then provided }$f(\mathbf{x%
},t;Z)$ \emph{is a function of} $(\mathbf{u,}p,\mathbf{r,}t)$
\emph{which depends slowly both on the relative velocity}
$\mathbf{u=v-V}$ \emph{and fluid pressure} $p,$ \emph{in the sense
that, in validity of the asymptotic orderings (\ref{finite
amplitude}) and (\ref{finite-amplitude 2}), the local
pdf is taken of the form }$f=f(\zeta \mathbf{u,}\zeta p,\mathbf{r,}t)$ \emph{%
(smoothness assumption). Then in follows that: }

\emph{B1) Eq.(\ref{Fokker-Planck operator}) holds also for a generic pdf\ }$%
\left\langle f\right\rangle ;$

\emph{B2) there exists a minimal representation for the F-P operator (\ref%
{Fokker-Planck operator}) obtained by retaining only the following
F-P coefficients:} $\mathbf{C}_{1,1}=-\left\langle \delta
\mathbf{F}\delta
p\right\rangle $ \emph{and} $\mathbf{C}_{i,0}=\left\langle \delta \mathbf{F}%
\delta A_{i}\right\rangle $ \emph{with} $i=2,3,4$ \emph{where respectively} $%
\delta A_{2}=\delta \mathbf{V,}\delta A_{3}=-\delta \mathbf{V}\delta \mathbf{%
V}$ \emph{and} $\delta A_{4}=\frac{1}{2}\delta
\mathbf{V}\left\langle \delta
\mathbf{V}\delta \mathbf{V}\right\rangle -\left[ \frac{1}{6}\delta \mathbf{V}%
\delta \mathbf{V}\delta \mathbf{V-}\left\langle \delta
\mathbf{V}\delta \mathbf{V}\delta \mathbf{V}\right\rangle \right]
,$ \emph{such that the stochastic-averaged kinetic equation Eqs.
(\ref{stochasrtic-averaged IKE})
recovers the exact stochastic-averaged fluid equations (\ref{average NS})-(%
\ref{physical realiz.2-average}).}\

PROOF - The proof (of proposition B$_{1}$) follows by explicitly evaluating $%
\delta \Delta f$ in terms of $f(\mathbf{x},t;\left\langle
Z\right\rangle ).$ This can be achieved formally by introducing a
suitable Taylor expansion for $\delta \Delta f$ in the
neighborhood of $f(\mathbf{x},t;\left\langle Z\right\rangle ).$
This is permitted if $f$ satisfies the smoothness assumption
(\ref{physical realiz.2-average}) indicated above, which implies
in particular that the orderings (\ref{finite amplitude}), (\ref%
{finite-amplitude 2}) must be satisfied. This requirement is
manifestly
satisfied, in particular, by $f\equiv f_{M}.$ For the general case in which $%
f\neq f_{M}$ this permits to estimate \ $\delta \Delta f$ in terms of \ $f(%
\mathbf{x},t;\left\langle Z\right\rangle )$ by a perturbative
(Taylor) expansion. In particular, to leading-order in $O(\zeta
),$\ $\delta \Delta f$ \ reads $\delta \Delta f\cong -\left(
\delta \mathbf{V\cdot }\frac{\partial
}{\partial \mathbf{v}}+\delta p\frac{\partial }{\partial P_{0}}\right) f(%
\mathbf{x},t;\left\langle Z\right\rangle )\left[ 1+o(\zeta
)\right] .$ Since the $n$-th order terms in Eq.(\ref{Fokker-Planck
operator}) result by construction of order $o(\zeta ^{n}),$ this
implies the asymptotic convergence of the expansion for $\delta f$
in the velocity sub-domain in which the ordering
(\ref{finite-amplitude 2}) holds. Furthermore, one can
prove that in order that the stochastic kinetic equations (\ref%
{stochasrtic-averaged IKE}) and (\ref{stochastic IKE}) yield a
consistent inverse kinetic theory for the stochastic-averaged N-S
equations (including the corresponding energy equation), it is
sufficient to retain only a finite number of F-P coefficients (as
indicated in proposition B$_{2}$). \ The proof by explicit
calculation of the relevant moment equations of IKE in which the
truncated F-P collision operator is invoked.

A basic consequence of THM.2 is that in the velocity-space
sub-domain\ defined by the inequality $\left\vert X\right\vert
\sim o(\zeta ),$ the stochastic-averaged kinetic equation
(\ref{average NS}) can be approximated
by a time and spatially-dependent F-P equation containing generally \emph{%
both velocity and pressure perturbations. \ }In particular, due to
the asymptotic ordering (\ref{finite amplitude}),
(\ref{finite-amplitude 2}), to leading order in $o(\zeta )$
pressure perturbations appear only through the explicit
contributions carried by $\delta f$ and $\mathbf{C}_{1,1}.$
Finally, the form of the F-P operator is independent of the
specific choice of the pdf, provided the above smoothness
assumptions are satisfied.

\section{Conclusions}

In this paper a statistical model of hydrodynamic turbulence has
been formulated, based on the IKT approach earlier developed by
Ellero et al. \cite{Ellero2000,
Ellero2005,Tessarotto2006,Tessarotto2008}, which holds for a
generic form of the stochastic probability density $g(\alpha )$.
Basic feature of the new theory is that it satisfies exactly the
full set of stochastic fluid equations while permitting, at the
same time, the
construction of the stochastic pdf which - in difference with Ref.\cite%
{Tessarotto2008-aa} - advances in time the full set of stochastic
fluid fields. Unlike customary statistical approaches, this is
identified with the local position-velocity joint probability
density function, rather than the two-point correlation function
(velocity-difference pdf). \ The present theory displays several
interesting features. In particular, the stochastic-averaged pdf
has been shown to satisfy an H-theorem, assuring its strict
positivity. The corresponding kinetic equation is formally similar
to the Vlasov equation arising in the strong turbulence theory of
Vlasov-Poisson plasmas \cite{Dupree1966,Weinstock1969}, for which
a renormalized kinetic theory can be in principle applied
\cite{Krommes1979}. \ However, unlike plasmas, the stochastic IKT
here considered admits exact local kinetic equilibria for
$\left\langle f(\mathbf{x},t;Z)\right\rangle $,
which are expressed by the stochastic-averaged Maxwellian distribution $%
\left\langle f_{M}(\mathbf{x},t;Z)\right\rangle $. \ More
generally, for pdf's which are suitably smooth, in the sense of
Thm.2, the stochastic-averaged kinetic equation can be
approximated in terms of Fokker-Planck kinetic equation. This
result suggests a possible new interesting viewpoint for the
investigation of turbulence theory in neutral fluids, which
includes, in particular, the analysis of previous F-P models of
turbulence \cite{Naert1997}. This topic will be the subject of a
forthcoming investigation.

\section*{Acknowledgments}
Work developed (M.T.) in the framework of the MIUR (Italian
Ministry of University and Research) PRIN Research Program
``Modelli della teoria cinetica matematica nello studio dei
sistemi complessi nelle scienze applicate'' and the European COST
action P17 (M.T). The partial of the GNFM (National Group of
Mathematical Physics) of INDAM (National Institute of Advanced
Mathematics, Italy) (M.T. and P.N.) and of the Deutsche
Forschungsgemeinschaft via the project EL503/1-1 (M.E.) is
acknowledged.

\section*{Notice}
$^{\S }$ contributed paper at RGD26 (Kyoto, Japan, July 2008).



\end{document}